\documentclass[aps,prx,reprint,showkeys]{revtex4-2}

\usepackage{lipsum}
\usepackage{blindtext}
\usepackage[export]{adjustbox}
\usepackage{graphicx,epsfig,epstopdf}
\usepackage{amsmath,amssymb}
\usepackage{esvect}
\usepackage{color}
\usepackage{subfigure}
\usepackage{array}
\usepackage{tabularx}
\usepackage{multirow}
\usepackage{booktabs}
\usepackage{mathtools}
\usepackage{float}
\usepackage{hyperref}
\newcolumntype{L}[1]{>{\raggedright\arraybackslash}p{#1}}
\newcolumntype{C}[1]{>{\centering\arraybackslash}p{#1}}
\newcolumntype{M}[1]{>{\centering\arraybackslash}m{#1}}
\allowdisplaybreaks

\def\_#1{{\bf #1}}

\def\.{\cdot}

\def\l#1{\label{eq:#1}}
\def\r#1{(\ref{eq:#1})}

\def\e{\begin{equation}}
\def\f{\end{equation}}
\def\*{^{\displaystyle*}}
\def\=#1{\overline{\overline #1}}

\def\.{\cdot}

\def\##1{{\bf #1\mit}}
\def\_#1{{\bf #1\mit}}
\def\-#1{{\bf #1\mit}}

\usepackage[version=4]{mhchem}

\def\am{\left(\begin{array}{c}}
\def\amm{\left(\begin{array}{cc}}
\def\a{\end{array}\right)}

\def\eeco{\alpha_{\rm e}}
\def\eecr{\alpha_{\rm ge}}
\def\mmco{\alpha_{\rm m}}
\def\mmcr{\alpha_{\rm gm}}
\def\te{\alpha_{\rm te}}

\def\om{\alpha_{\rm om}}

\def\=#1{\overline{\overline #1}}

\begin{document}

\title{Realization of the Tellegen Effect in Resonant Optical Metasurfaces}

% \title{Optical Tellegen Nonreciprocal Metasurface}

%\title{Realization of optical Tellegen metasurface}

\author{Shadi~Safaei Jazi$^{1,*}$} %\email{Corresponding author: shadi.safaeijazi@aalto.fi} 
\author{Ihar~Faniayeu$^{2,*}$}
\author{Rafael~Cichelero$^{2}$}
\author{Nikolai~Kuznetsov$^{3}$}
\author{Sebastiaan~van~Dijken$^{3}$}
\author{Shanhui~Fan$^{4}$}
\author{Alexandre~Dmitriev$^{2,*}$}
\author{Viktar~Asadchy$^{1}$} 
\email{Corresponding authors: shadi.safaeijazi@aalto.fi,\\ ihar.faniayeu@physics.gu.se, alexd@physics.gu.se,\\  viktar.asadchy@aalto.fi}

\affiliation{
$^1$Department of Electronics and Nanoengineering, Aalto University, P.O.~Box 15500, FI-00076 Aalto, Finland\\
$^2$Department of Physics, University of Gothenburg, Gothenburg 41296, Sweden\\
$^3$Department of Applied Physics, Aalto University, P.O.~Box 15100, FI-00076 Aalto, Finland\\
$^4$Ginzton Laboratory and Department of Electrical Engineering, Stanford University, Stanford, California 94305, USA
}

\begin{abstract}
\section*{Abstract}
The nonreciprocal magnetoelectric effect in Tellegen materials enables exotic phenomena such as axion-modified electrodynamics and fosters the development of magnet-free nonreciprocal media. As the nonreciprocal counterpart to the well-known chiral electromagnetic response, it offers a parallel framework in which many concepts developed for chiral materials can be translated to Tellegen media, potentially unlocking new avenues for fundamental studies and applications. Although predicted over 75 years ago and observed in only a handful of natural materials with very low strength, the strong optical Tellegen effect has remained experimentally elusive. Here, we report the first experimental demonstration of a resonant optical diagonal Tellegen effect in a metasurface, showcasing a response that is 100 times greater than that of any known natural material. This optical metasurface, consisting of randomly distributed cobalt-silicon nanoscatterers with strong shape anisotropy, utilizes spontaneous magnetization to achieve a robust Tellegen effect without the need for an external magnetic field. In addition to the Tellegen response, the metasurface exhibits both gyroelectric and gyromagnetic effects, contributing to nonreciprocal cross-polarized light reflection. We introduce a technique to independently extract the amplitudes of these three effects using conventional magneto-optical single-side-illumination measurements. The observation of the resonant Tellegen effects in the optical frequency range may lead to the experimental observation of axionic electrodynamics and compact bias-free nonreciprocal optical devices.
% ~\cite{fiebig_revival_2005, sikivie_experimental_1983,wilczek_two_1987, PhysRevLett.Marsh2019, NatRevPhys.Nenno2020,PhysRevLett.Essin2009,lindell_electromagnetic_1994,Nat.Commun.Safaei2024}.
\end{abstract}

\keywords{}

\maketitle

\section{Introduction}
% Linear materials interact with electromagnetic waves via permittivity and permeability that describe the interaction by electric and magnetic polarizations, while magnetoelectric coupling parameters can also be added ~\cite{serdyukov_electromagnetics_2001}. 
Magnetoelectric coupling is a defining characteristic of bianisotropic materials, where a magnetic field induces electric polarization and, conversely, an electric field induces magnetization. Bianisotropic phenomena are categorized into reciprocal (commonly known as chirality) and nonreciprocal (Tellegen) effects~\cite{serdyukov_electromagnetics_2001,asadchy_bianisotropic_2018}. Chirality, which arises in materials that exhibit reciprocity but lack parity, has been extensively studied in natural and engineered materials ~\cite{caloz_electromagnetic_2020,Solomon_acsphotonics,Nat.Chen2023,Xudong_Science}. In contrast, the nonreciprocal magnetoelectric effect, named after Bernard Tellegen, who first proposed it in 1948~\cite{Tellegen1948}, is extremely rare as it requires the simultaneous breaking of both parity symmetry and reciprocity~\cite{hill_why_2000,asadchy_tutorial_2020}.

% for practical applications, while it is expected to have a groundbreaking impact on both fundamental and applied physics~\cite{fiebig_revival_2005}.

The diagonal Tellegen effect is expected to have a significant impact on both fundamental and applied physics~\cite{fiebig_revival_2005}. It should be clearly distinguished from its off-diagonal counterpart, the magnetochiral effect~\cite{rikken_observation_1997-2,train_strong_2008}, as it necessitates a diagonal magnetoelectric tensor with a non-zero trace.   
Seminal works have demonstrated that the electrodynamic equations governing isotropic Tellegen materials take the same form as those describing a hypothetical axion medium within the framework of quantum field theory~\cite{sikivie_experimental_1983,wilczek_two_1987}. The axion is an elementary particle theorized to address two major issues in modern physics: the strong charge conjugation and parity (CP) problem~\cite{peccei_cp_1977} and the existence of dark matter~\cite{abbott_cosmological_1983}.
Tellegen-type nonreciprocal materials have attracted considerable interest as potential enablers of exotic concepts of axion electrodynamics in practical materials, such as dyon quasiparticles, anyon statistics, the Witten effect, and axionic polaritons~\cite{wilczek_two_1987,NatPhys.Li2010,PhysRevLett.Tercas2018,PhysRevB.Barredo-Alamilla2024}. They may also play a crucial role in the search for axions~\cite{PhysRevLett.Marsh2019, NatRevPhys.Nenno2020, PhysRevLett.Essin2009}.
Moreover, Tellegen materials promise a range of exotic phenomena, especially in the optical regime, including nonreciprocal light reflection~\cite{lindell_electromagnetic_1994,Nat.Commun.Safaei2024} and transmission~\cite{prudencio_optical_2016}, photonic topological edge states~\cite{he_photonic_2016}, PT-symmetric magnetoelectric energy density~\cite{bliokh_magnetoelectric_2014},   and spin-dependent thermal radiation~\cite{khandekar_new_2020}.
% Both isotropic and anisotropic Tellegen materials promise a range of exotic phenomena, especially in the optical regime, including nonreciprocal light reflection~\cite{lindell_electromagnetic_1994,Nat.Commun.Safaei2024} and transmission~\cite{prudencio_optical_2016}, photonic topological edge states~\cite{he_photonic_2016}, PT-symmetric magnetoelectric energy density~\cite{bliokh_magnetoelectric_2014}, synthetic movement~\cite{caloz_spacetime_2020}, directional dichroism~\cite{rikken_observation_1997-2,train_strong_2008}, and spin-dependent thermal radiation~\cite{khandekar_new_2020}.

While the diagonal Tellegen effect could be measured at microwave frequencies for certain multiferroic materials~\cite{fiebig_revival_2005,eerenstein_multiferroic_2006,Pyatakov2012}, it already becomes two orders of magnitude smaller than the refractive index in the terahertz range~\cite{PhysRevLett.Kurumaji2017} and five orders of magnitude smaller in the optical regime~\cite{Rivera1994, 
PhysRevLett.Krichevtsov1996}. This drastic reduction at optical frequencies is attributed to the fact that both cyclotron and Larmor frequencies of electrons responsible for breaking the Lorentz reciprocity typically lie in the gigahertz range. 
Thus, artificial Tellegen composites designs have been confined to the microwave frequencies: from individual constituents (meta-atoms)~\cite{kamenetskii_experimental_2000, tretyakov_artificial_2003, mirmoosa_polarizabilities_2014} to the two-dimensional arrays (metasurfaces)~\cite{kamenetskii_technology_1996,PhysRevB.Radi2016,NatCommun.Yang2025}. 
The quest for optical Tellegen effect has inspired three recent conceptual proposals for Tellegen metamaterials. The first two approaches, based on rapid temporal modulation of material properties~\cite{PhysRevApplied.19.024031,Shaposhnikov2024} and multilayer antiferromagnetic configuration~\cite{PhysRevB.108.115101,Seidov2024}, are not achievable with the current manufacturing technology. The third theoretical proposal based on metasurfaces with spontaneous magnetization in magnetic two-component nanoparticles has not been experimentally validated either~\cite{Nat.Commun.Safaei2024}.

% In the optical spectrum, the observed ME phenomenon in known materials is negligible due to the insignificant magnetization effects of materials~\cite{lindell_electromagnetic_1994}; nevertheless, it can be amplified by leveraging the metamaterial paradigm~\cite{xiao_loss-free_2010}. In contrast, the ME effect is substantial at microwave frequencies because of strong magnetic responses~\cite{eerenstein_multiferroic_2006}. 
% Over the past two decades, several approaches have been explored to attain artificial ME composites based on the principles of metamaterials—surpassing that observed in certain natural materials. However, the meta-atoms designed to date have been primarily limited to microwave frequencies~\cite{kamenetskii_technology_1996,kamenetskii_experimental_2000,tretyakov_artificial_2003,mirmoosa_polarizabilities_2014,yang2024tellegenresponsesmetamaterials}. 
% Furthermore, achieving the ME effect in the proposed meta-atoms requires a uniform external magnetic bias. However, the isotropic ME effect in three-dimensional metamaterials composed of such meta-atoms disappears when a uniform external magnetic bias is applied. 
 
Here, for the first time, we fabricate and experimentally characterize an optical Tellegen metasurface with a diagonal magnetoelectric tensor.
The metasurface consists of nanoscale cone-shaped scatterers, composed of amorphous silicon and cobalt. Due to strong shape anisotropy, the cobalt part is in a single-domain magnetic state at fabrication, ensuring that the metasurface displays the Tellegen effect even in the absence of external magnetization. We further demonstrate that the optical Tellegen metasurface also exhibits significant gyroelectric and nontrivial gyromagnetic~\cite{yang_observation_2022} responses.
As all three effects collectively contribute to the non-reciprocal reflection of the cross-polarized light from the metasurface, we propose an original method to independently determine their respective amplitudes using conventional magneto-optical measurements. The metasurface displays a Tellegen parameter on the order of $10^{-3}$, which aligns well with theoretical and simulation predictions and is at least 100 times larger than values observed in natural materials such as chromium oxide~\cite{B.B.Krichevtsov_1993}.
Moreover, the chosen nanoscatterers geometry enables producing large-scale optical Tellegen metasurfaces using high-throughput bottom-up nanofabrication, allowing straightforward scalability to hundreds of square centimeters.

% Lately, the first optical Tellegen metamaterial without requiring external magnetic bias and utilizing conventional materials and nanofabrication techniques has been theoretically proposed~\cite{Nat.Commun.Safaei2024}. Although it was the first realistic proposal, it is inappropriate for large-scale fabrication. Inspired by the proposed metamaterial, we fabricate and experimentally characterize the first optical Tellegen metasurface with an effective method to enable large-scale fabrication, in this study. 
% We show that the Tellegen response, which is responsible for isotropic Kerr rotation, in addition to magnetic and electric gyrotropic responses contributes to the optical Kerr rotation observed in the metasurface. We develop a novel approach to distinguish the Tellegen response of such metasurfaces from the gyrotropic effects. As the metasurface can be fabricated over a large scale, it provides an initial step toward the development of fully three-dimensional colloidal structures.

\section{Results}
\subsection{Metasurface design}\label{design}

The isotropic Tellegen effect is manifested by a nonreciprocal cross-polarized light reflection at a vacuum-matter interface. For spatially uniform slabs, the effect remains undetectable in light transmission~\cite{lindell_electromagnetic_1994}. Unlike magneto-optical effects of both gyroelectric and gyromagnetic types, Tellegen media exhibit a direction-independent Kerr effect due to isotropy and do not exhibit the Faraday effect~\cite{Nat.Commun.Safaei2024}. This distinction makes measuring the Tellegen effect more challenging, as its observable manifestation—the ``isotropic" Kerr effect—depends solely on the concentration of Tellegen meta-atoms rather than their total number or the slab thickness~\cite[p.~261]{lindell_electromagnetic_1994}.
Therefore, the fabrication and experimental verification of high-density optical Tellegen metasurfaces, where interface phenomena are readily observable, represent the most straightforward approach to exploring the Tellegen effect.
% Moreover, the fabrication of such metasurfaces is a crucial step toward developing bulk isotropic Tellegen colloids by dispersing meta-atoms into a liquid solution via a lift-off procedure. Importantly, achieving a high volume concentration of meta-atoms in such colloids requires the fabrication of metasurfaces with large footprints. To this end, we employ hole-mask colloidal lithography (HCL), a technique well-suited for producing large-scale, cost-effective metasurfaces~\cite{HCL}.

\begin{figure*}[tb]
	\centering
	\includegraphics[width=1\linewidth]{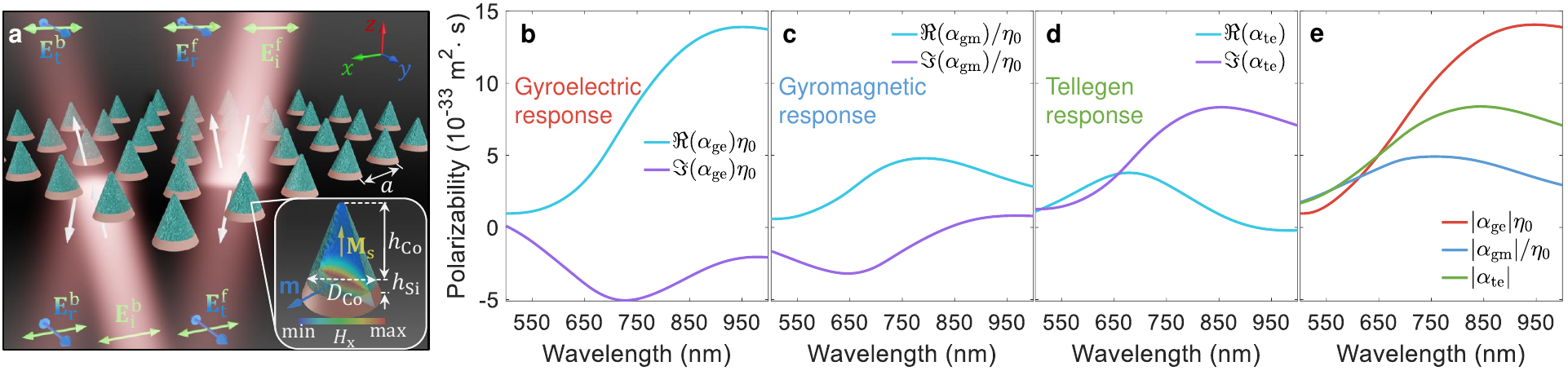} 
	\caption{\textbf{Optical Tellegen metasurface design.} (a) Illustration of the metasurface with optimized optical Tellegen meta-atoms consisting of cobalt (top) and silicon (bottom) parts. The metasurface is illuminated from opposite directions by normally incident light (a slight oblique incidence is depicted for visual clarity). The gradient color indicates that the reflected and transmitted light contain both co- and cross-polarized components. The inset depicts a close-up image of the meta-atom with close to single-domain vertical magnetization $\_M_{\rm s}$  in the cobalt. It also shows the high-frequency magnetic field distribution inside the nanocone when illuminated by a standing wave with an electric field antinode positioned at the center of the meta-atom. The field distribution is calculated at a free-space wavelength of 830~nm and reveals an $x$-oriented magnetic dipole moment $\_m$ induced solely by the incident $x$-polarized electric field through the isotropic (diagonal) Tellegen effect.
     (b)--(e) Simulated complex polarizabilities of the metasurface in (a), normalized by the free-space impedance $\eta_0$.  
    }
	\label{fig1}
\end{figure*}

We select the Tellegen meta-atom as a nanoscale cone-shaped scatterer, shown in the inset of Fig.~\ref{fig1}(a). The upper part of the nanocone is cobalt, with a height \( h_{\rm Co} = 119 \)~nm and a base diameter \( D_{\rm Co} = 76 \)~nm. The lower part is amorphous silicon, with a height \( h_{\rm Si} = 20 \)~nm and a base diameter \( D_{\rm Si} = 89 \)~nm. This  geometry has an apex angle of approximately \( 35^\circ \), enabling straightforward fabrication via bottom-up hole-mask colloidal lithography (HCL). Cobalt possesses one of the largest magneto-optical parameters in the visible spectrum~\cite{vanengen_magnetic_1983}. The optimal base diameter of the cobalt part, \( D_{\rm Co} \), was determined by fabricating test metasurfaces consisting of cobalt-only nanocones to achieve a configuration closest to the ideal single-domain state (see Supplementary Section~1) that arises due to the strong shape anisotropy~\cite{coey_magnetism_2010-1} and remains unaffected by the addition of the silicon. That is, the cobalt part can sustain strong vertical remanent magnetization close to the saturation magnetization \(  M_{\rm s}\)  in the absence of an external magnetic field. The optimal height of the silicon part was determined by full-wave simulations to induce a magnetic-type Mie resonance~\cite{Nat.Commun.Safaei2024}, thereby enhancing the Tellegen response of the meta-atom (see  Supplementary Section~2). The magnetic field distribution of this resonant mode inside the nanocone is depicted in the inset of Fig.~\ref{fig1}(a) at a free-space wavelength of 830~nm. Here, the excitation consists of a standing wave with an  $x$-polarized electric field antinode located at the center of the meta-atom. This field alone induces an $x$-oriented magnetic dipole moment $\mathbf{m}$ inside the nanocone. This behavior is a direct manifestation of the Tellegen effect~\cite{asadchy_bianisotropic_2018}. Although the chiral magnetoelectric effect could similarly give rise to a magnetic dipole excitation, it vanishes due to the symmetry of the nanocone, as discussed below.

% In other words, magnetic dipole moments in optics come from circulating currents driven by the optical field (spatial dispersion effect). A circulating conducting or displacement current generates an axial magnetic field hotspot ($H_x$ in this case).

The nanocone possesses the magnetic point group \( C_{\infty v} (C_{\infty}) \), indicating that it can support only gyroelectric, gyromagnetic, reciprocal bianisotropic (omega coupling), and nonreciprocal (diagonal) Tellegen effects. Here, we refer to gyroelectric and gyromagnetic effects as those responsible for Kerr and Faraday rotation in materials with off-diagonal permittivity and permeability tensors, respectively.
Notably, the symmetry of the meta-atom inherently prohibits any chirality. A detailed analysis of the electromagnetic material tensors corresponding to this point group can be found in~\cite{Nat.Commun.Safaei2024}.
It is important to highlight that the chosen meta-atom exhibits both parity and reciprocity breaking, leading to the emergence of the Tellegen effect even in the absence of silicon Mie resonator part. However, adding it results in a two-fold enhancement of the Tellegen effect due to the resonance mechanism (see Supplementary Section~2).

We first consider the metasurface in free space with meta-atoms  in a square periodic lattice (Fig.~\ref{fig1}(a).) We analyze the metasurface response using full-wave simulations and extract the collective polarizabilities, which relate the induced electric and magnetic dipole moments (\(\mathbf{p}\) and \(\mathbf{m}\)) to the incident electric and magnetic fields (\(\mathbf{E}\) and \(\mathbf{H}\))~\cite{asadchy_bianisotropic_2018}:
\begin{equation}
\label{eq:dipoles}
\begin{array}{c}\displaystyle
    \_p = \left[ \eeco \overline{\overline{I}} + \eecr \overline{\overline{J}} \right] \cdot \_E + 
          \left[  \te    \overline{\overline{I}} +  \om    \overline{\overline{J}}   \right] \cdot  \_H, \vspace{6pt} 
    \\ \displaystyle 
    \_m = \left[ \mmco \overline{\overline{I}} + \mmcr \overline{\overline{J}}  \right] \cdot \_H + 
          \left[  \te    \overline{\overline{I}} +  \om   \overline{\overline{J}}   \right]\cdot  \_E,
    % \_p = \left[ \eeco I + \eecr J \right] \_E + 
    %       \left[ (\te  + \ch) I + (\om  + \mo) J  \right] \_H, \vspace{6pt} 
    % \\ \displaystyle 
    % \_m = \left[ \mmco I + \mmcr J \right] \_H + 
    %       \left[ (\te  - \ch) I + (\om  - \mo) J  \right] \_E.
    \end{array}
% \tag{1}
\end{equation}
where \( \overline{\overline{I}}  \) represents the two-dimensional identity matrix, and \( \overline{\overline{J}}  \) denotes the antisymmetric (skew-symmetric) two-dimensional matrix characterized by the nonzero components 
\( J_{yx} = -J_{xy} = 1 \). We limit our analysis to the dipolar moments since they provide a dominant contribution to the total scattering of the considered nanocone (see Supplementary Section~3). The quadrupole contribution is neglected as an approximation because it is at least one order of magnitude smaller than the corresponding dipolar contributions.
Thus, we model the nanocones as bianisotropic meta-atoms (with weak spatial dispersion~\cite{serdyukov_electromagnetics_2001}) described by the local material relations~(\ref{eq:dipoles}).
Light is incident on the metasurface along the \( z \)-axis (normal to the surface), which is also the uniaxial direction of the metasurface.
The collective polarizabilities \( \eeco \) and \( \mmco \) represent the ordinary reciprocal electric and magnetic responses, respectively. The terms \( \alpha_{\rm ge} \) and \( \alpha_{\rm gm} \) correspond to the nonreciprocal gyroelectric and gyromagnetic responses, while \( \te \) and \( \om \) describe the nonreciprocal Tellegen and reciprocal omega responses, respectively. All the polarizabilities are defined with respect to the base of the nanocone.
It should be noted that the collective polarizabilities of the metasurface in~(\ref{eq:dipoles}) depend on the angle of incidence, even when the meta-atoms are perfectly local, i.e., characterized by angle-independent individual polarizabilities. This dependence arises from the mutual interactions among the meta-atoms within the lattice, which constitute an angle-dependent phenomenon~\cite{7870611}. Nevertheless, under normal incidence and for sufficiently sparse metasurfaces, the collective and individual polarizabilities of a dipolar meta-atom are qualitatively similar and differ in magnitudes by a nearly constant small factor~\cite{niemi_polarization_2012, PhysRevB.89.075109}. Therefore, in the following, we describe the Tellegen metasurface using collective polarizabilities, as they are significantly easier to extract experimentally and have the same order of magnitude as the individual ones.

\begin{figure*}[tb]
	\centering
	\includegraphics[width=0.9\linewidth]{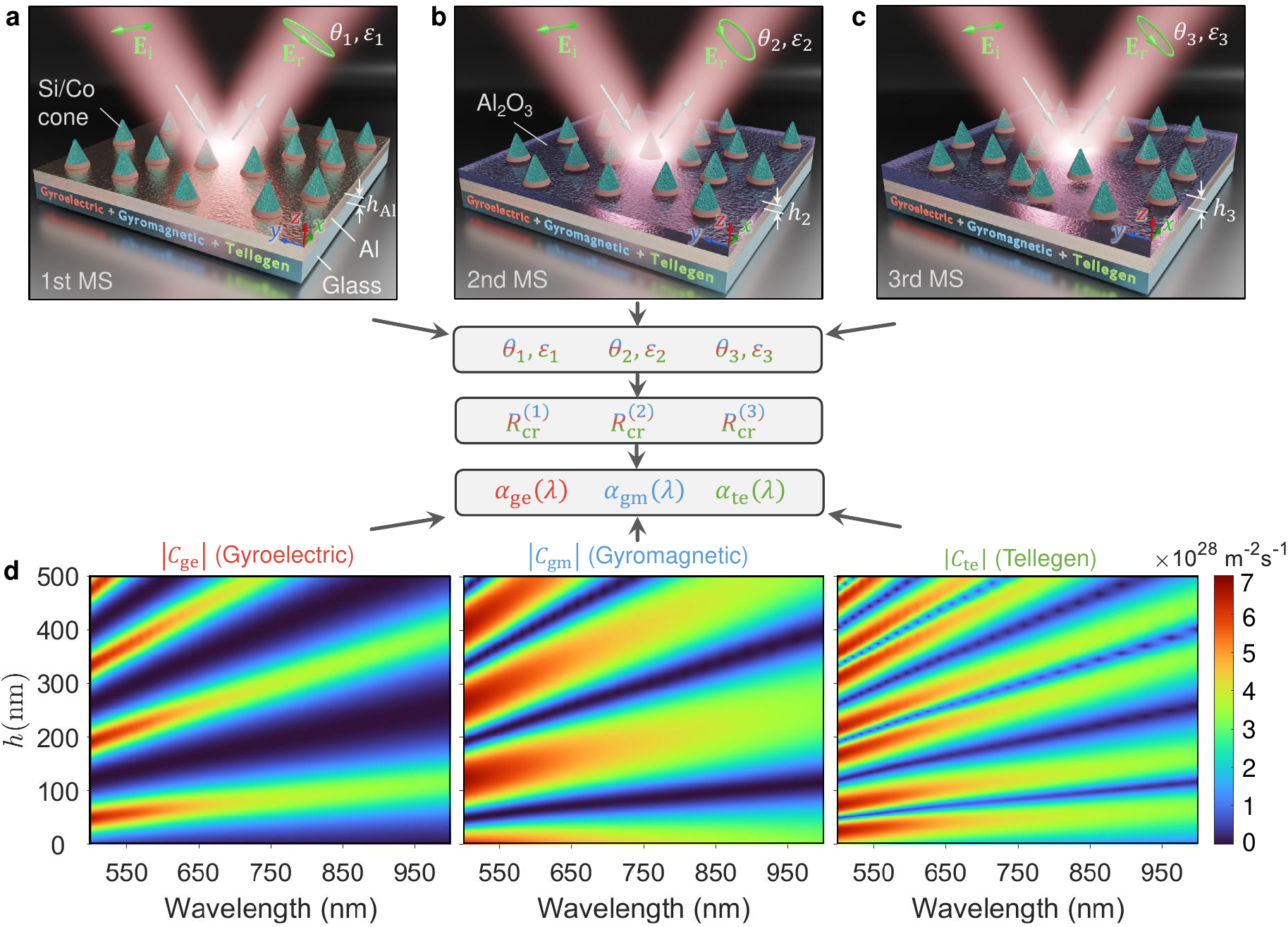} 
	\caption{\textbf{Extraction of the independent gyrotropic, gyromagnetic, and Tellegen effects in a general nonreciprocal metasurface using magneto-optical measurements.} (a)--(c) Three equivalent metasurfaces with identical meta-atoms and the same average surface densities on top of alumina spacers of three different thicknesses $h_1=0$~nm, $h_2=60$~nm, and $h_3=120$~nm. 
The metasurface are illuminated from the top by a normally incident light, and the Kerr rotation $\theta$ and ellipticity $\epsilon$ are extracted (an oblique incidence is depicted only for visual clarity). (d) 
    The complex amplitudes of the coefficients introduced in Eq.~\ref{eq:Rcr} plotted as functions of the free-space wavelength $\lambda$ and the alumina spacer thickness $h$. These coefficients quantify the contributions of the gyroelectric, gyromagnetic, and Tellegen effects to cross-polarized light reflection. 
    % The vertical dashed lines indicate the wavelengths at which the Tellegen effect is resonant. The horizontal dashed lines mark the selected alumina thicknesses $h_{\rm ge}$, $h_{\rm te}$, and $h_{\rm gm}$, corresponding to the scenarios where the contributions from each distinct effect are maximized. 
    }
	\label{fig2}
\end{figure*}

The Tellegen polarizability $\te$ can be calculated for the considered metasurface as
\begin{equation}
    \te = \frac{i a^2 }{2 \omega} (R_{\rm f}^{\rm cr}-R_{\rm b}^{\rm cr}),
    \l{ate}
\end{equation}
where $R_{\rm f}^{\rm cr}$ and $R_{\rm b}^{\rm cr}$ denote complex-amplitude cross-polarized reflection coefficients for the forward and backward illuminations, respectively, $a=320$~nm is the metasurface period, and $i$ is the imaginary unit (we adopt time-harmonic oscillations of the form $e^{i \omega t}$). This and other expressions for the unknown polarizabilities are provided in Supplementary Section~3. 
As observed in~\r{ate}, extracting the Tellegen effect in the metasurface requires illumination from both sides. This is due to the fact that all three effects—gyroelectric, gyromagnetic, and Tellegen—are nonreciprocal and collectively contribute to the cross-polarized light reflection. Consequently, to determine separately the Tellegen effect in a metasurface that may also display gyroelectric and/or gyromagnetic responses, one must evaluate the specific linear combination of the two complex reflection coefficients. This necessitates precise measurement of both their complex amplitudes and phases.

By evaluating the full scattering matrix of the metasurface through full-wave simulations, we compute the three nonreciprocal collective polarizabilities (see Supplementary Section~3), as shown in Figs.~\ref{fig1}(b)--(d). From Fig.~\ref{fig1}(e), it can be observed that all three polarizabilities exhibit similar magnitudes, necessitating illumination from both sides for their accurate extraction.
At approximately 830~nm, the Tellegen response exhibits a Lorentzian-type resonance. Additionally, the Tellegen meta-atom demonstrates a resonant gyromagnetic response, whose presence at optical frequencies was only recently observed in metamaterials~\cite{yang_observation_2022,Laser&PhotonicsReviews2022}. The origin of this gyromagnetic effect can be explained as the combination of weak spatial dispersion and the gyroelectric effect inside the nanocone (see Supplementary Section~3).

Crucially, when the metasurface is experimentally realized on a substrate, measuring the complex-valued reflection coefficients for both illumination directions becomes exceedingly challenging. Due to the finite bandwidths of supercontinuum lasers required for broadband measurements, the large thickness of a typical substrate would exceed the coherence length, driving the system into an incoherent regime. This would make phase measurements of the reflection coefficients for both forward and backward illuminations in practice impossible. The substantial total thickness variation of the substrate, caused by non-uniform roughness, would further complicate phase measurements, as determining the thickness of such a thick substrate with high precision is challenging. Without phase information, the Tellegen response cannot be accurately extracted, as the gyroeletric and gyromagnetic effects would contribute to the measurements, as discussed above.

\subsection{Experimental extraction of the Tellegen effect}\label{extraction}

To address these challenges, we introduce a novel
method: we fabricate and characterize \textit{three} Tellegen metasurfaces, measuring their reflection responses under illumination from a \textit{single} side. These metasurfaces are structurally identical, each composed of the same type of meta-atoms arranged with an equal average surface density, as illustrated in Figs.~\ref{fig2}(a)--(c). The key distinction lies in the thickness of the dielectric spacers on which they are mounted. Beneath each dielectric spacer is a continuous 100~nm-thick aluminum layer, followed by a glass (\(\text{SiO}_2\)) substrate with dimensions \(24 \, \text{mm} \times 24 \, \text{mm} \times 0.4 \, \text{mm}\), which ensures nearly complete reflection of incident light.
The dielectric spacer is composed of alumina (\(\text{Al}_2\text{O}_3\)), chosen for its low optical loss. The thin and precisely controlled thickness of alumina spacer enables straightforward and reliable optical measurements of the complex reflection coefficient.
As stated above, the cross-polarized reflection is influenced not only by the Tellegen effect but also by gyroelectric and gyromagnetic effects in the metasurface. To independently determine the amplitudes of these three effects, measurements from three metasurfaces are required.

For the metasurfaces considered in Figs.~\ref{fig2}(a)--(c), the cross-polarized reflection of normally incident light linearly depends on the three polarizabilities according to (see Supplementary Section~4):
\begin{equation}
\label{eq:Rcr}
\begin{array}{c}\displaystyle
    R_{\text{cr}} = C_{\rm ge} \, \eta_0\eecr   + C_{\rm gm} \,\mmcr/\eta_0 + C_{\rm te} \te,
\end{array}
% \tag{2}
\end{equation}
where we introduce \( C_{\rm ge} \), \( C_{\rm gm} \), and \( C_{\rm te} \) as the weight amplitudes of the gyroelectric, gyromagnetic, and Tellegen effects in the cross-polarized reflected light. These amplitudes depend on the free-space wavelength $\lambda$, periodicity of the metasurface $a$, as well as material properties and thicknesses of the dielectric spacer and metallic layer. Importantly, these coefficients are independent of the meta-atoms and represent only the environment surrounding the metasurface. The free-space impedance \( \eta_0 \)  is included in (\ref{eq:Rcr}) to ensure that the polarizabilities remain normalized and have consistent dimensions. In our analysis, we assume that substrate-induced bianisotropy effects~\cite{miroshnichenko2015substrate,sinev2016polarization,albooyeh2015revisiting} can be neglected, as confirmed by our full-wave simulations (see Supplementary Section~4).

The weight coefficients are plotted versus the free-space wavelength $\lambda$ and the dielectric spacer thickness $h$ in Fig.~\ref{fig2}(d). For a given value of $\lambda$ the three weight coefficients exhibit maxima and minima along the \( h \)-axis, with the periodicity given by $\Delta h=\lambda/(2n_{\rm d})$ (see Supplementary Section~4), where $ n_{\rm d}$ is the refractive index of the dielectric spacer.
We see that for some specific values of the spacer thickness $h$, 
the gyroelectric (or gyromagnetic) effect is nearly the sole contributor to the cross-polarized reflection \( R_{\mathrm{cr}} \). This implies that by selecting such spacer thicknesses, one can reliably isolate and measure the gyroelectric (or gyromagnetic) properties of the metasurface deposited on the spacer using a \textit{single} measurement. This holds true regardless of the possible presence of other nonreciprocal effects. However, such independent extraction of gyroelectric (or gyromagnetic) effects is effective only at a single wavelength. To determine these effects over a broad wavelength range, measurements must still be performed on three metasurfaces with different spacer thicknesses.
Further, a dielectric spacer at which \( R_{\mathrm{cr}} \) depends solely on the Tellegen effect of the metasurface (\(\te\)) even at a single wavelength is not achievable. As can be seen in Fig.~\ref{fig2}(d), no spacer thickness \( h \) exists for which both the gyroelectric and gyromagnetic contributions vanish simultaneously.
Consequently, it is impossible to measure the Tellegen effect from a general nonreciprocal metasurface using a single metasurface and single-side illumination. 

To determine the Tellegen effect of the proposed optical metasurface, we use the three-metasurface method illustrated in Fig.~\ref{fig2}. Three metasurfaces with dielectric spacer thicknesses $h_1=0$~nm, $h_2=60$~nm, and $h_3=120$~nm are fabricated, approximately corresponding to the maxima of the weight amplitudes within the wavelength range of interest (around 830~nm). For each metasurface $(j)$, the complex cross-polarized reflection coefficient is obtained from conventional magneto-optical Kerr measurements as $R_{\mathrm{cr}}^{(j)}=(\theta_j+i\epsilon_j)R_{\mathrm{co}}^{(j)}$, where $\theta_j$ and $\epsilon_j$ are the Kerr rotation and ellipticity, respectively, and $R_{\mathrm{co}}^{(j)}$ is the co-polarized complex reflection coefficient~\cite{asadchy_tutorial_2020}. The three measured $R_{\mathrm{cr}}^{(j)}$ values, together with the corresponding calculated weight amplitudes $C_{\mathrm{ge}}$, $C_{\mathrm{gm}}$, and $C_{\mathrm{te}}$, provide a system of linearly independent equations that can be solved for the unknown polarizabilities $\alpha_{\mathrm{ge}}$, $\alpha_{\mathrm{gm}}$, and $\alpha_{\mathrm{te}}$ (see Supplementary Section~4 for details).

\subsection{Experimental realization of the optical Tellegen metasurface}\label{fabrication}
\begin{figure*}[tb]
	\centering
	\includegraphics[width=1\linewidth]{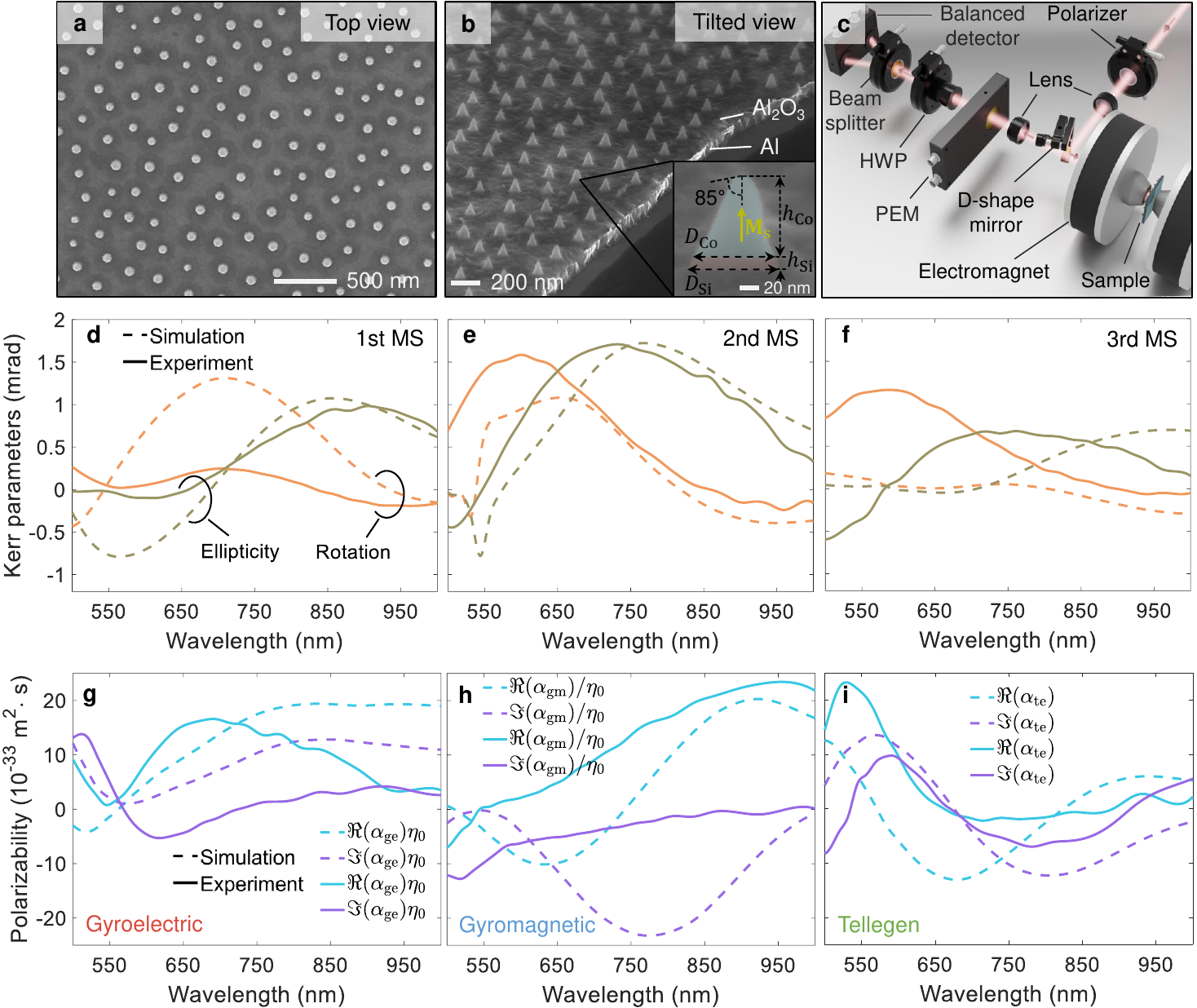} 
	\caption{\textbf{Fabricated Tellegen metasurface and its characterization.} (a)--(b) SEM images of one of the fabricated Tellegen metasurfaces with dielectric spacer thickness of $h=60$~nm (top and tilted views). The inset in (b) presents a zoomed-in image of a single meta-atom, artificially colored to visualize the cobalt and silicon parts with the averaged geometrical parameters of the nanocone used in full-wave simulations:  $h_{\rm Si}=20$~nm, $h_{\rm Co}=104$~nm, $D_{\rm Si}=90$~nm, and $D_{\rm Co}=84$~nm. The averaged half-apex angle of the nanocone is $85^\circ$, which reflects the degree of tip rounding due to the hole-mask colloidal lithography process.  (c) Schematic illustration of the polar magneto-optic Kerr effect (MOKE) experimental setup. Detailed information about the setup is available in the Methods section. (d)--(f) Experimental and simulated magneto-optical Kerr spectra for the three fabricated metasurfaces under an external magnetic field fixed at $B = 1.3$~T. Here, the positive-value magnetic field denotes the direction from the base of the meta-atom to its apex. The simulated spectra correspond to periodic metasurfaces with geometric parameters averaged from the fabricated samples. (g)--(i) The gyroelectric, gyromagnetic, and Tellegen polarizabilities extracted from the data in (d)--(f).
    }
	\label{fig3}
\end{figure*}

Three metasurfaces were fabricated using hole-mask colloidal lithography (see Materials and Methods Section). 
Figures~\ref{fig3}(a) and (b) present scanning electron microscopy (SEM) images of one of the fabricated metasurfaces.
% The structure of the fabricated metasurface, consisting of conical meta-atoms with a base diameter of 90 nm and a height of 140 nm, 
% was characterized using scanning electron microscopy (SEM) and atomic force microscopy (AFM). 
% Notably, the metasurface covers an area of $2 \times 2$ $ \text{cm}^2 $; however, it can be easily scaled up to $15 \times 15$ $ \text{cm}^2 $ because of the large scalability of the HCL method. 
The images confirm the conical shape of the meta-atoms, according to the design, while meta-atoms distribution provides a quasi-uniform surface density. 
% Figures~\ref{fig3}(d) and (e) ...
By extracting meta-atoms average surface density using the SEM images, we find the mean interparticle distance to be $a=295$~nm. The calculated mean base diameter of the meta-atoms is $D_{\rm Si}=90$~nm with a standard deviation of 23~nm.  
% The images further reveal that the metasurface consists of a quasi-periodic array with an average periodicity of 300 nm, which deviates from the idealized scenario where the array of meta-atoms was assumed to be periodically arranged with a periodicity of 319 nm. It is important to emphasize that the average surface density of the conical meta-atoms within the array is a critical factor influencing the performance of the metasurface.
The average height of the cones $h_{\rm Co} + h_{\rm Si}$ is about 130~nm, which is slightly lower than the 140~nm expected for an idealized nanocone. However, the cones exhibit rounded apexes rather than sharp ones. This characteristic is also evident in the SEM images (inset in Fig.~\ref{fig3}(b)) and AFM images (see Supplementary Section~5), where the apex appears circular rather than pointed. 
% To ensure a fair comparison between experimental and simulated results, our full-wave simulations feature meta-atoms with a rounded apex. 
To
ensure a fair comparison between experimental and simulated
results, we modified the meta-atom topology in our
full-wave simulations by replacing the idealized nanocone
with a sharp apex with one featuring a rounded apex.
An inset in Fig.~\ref{fig3}(b) illustrates the geometric parameters of the simulated nanocone specified in the figure caption. 
% By optimizing this refined model, we performed full-wave simulations to examine the Kerr effect in the three configurations, along with the Tellegen and gyrotropic responses of the metasurface.

The fabricated nanocones are generally symmetric, as can be seen in the top-view Fig.~\ref{fig3}(a) (dark regions centered on the bright spots). Although some imperfections and slight tilts can be present in the cone geometry (Fig.~\ref{fig3}(b)), they do not have preferred directions, and their statistical averaging would result in zero net chirality.  
Importantly, even if weak chirality were locally introduced due to such imperfections, it would not contribute to nonreciprocal optical rotation in our configuration. Indeed, a chiral layer placed on top of a mirror does not produce a net polarization rotation, as the rotation accumulated during forward propagation is compensated by the rotation accumulated upon reflection during backward propagation. Therefore, any residual chirality in the fabricated structures does not affect the observed Tellegen effect. In sharp contrast, the polarization rotation due to the Tellegen effect doubles after forward and backward propagations.

We measure the magneto-optical Kerr rotation angle ($\theta$) and Kerr ellipticity ($\epsilon$) using a magneto-optical setup illustrated in Fig.~\ref{fig3}(c) (see more details in Methods and Supplementary Section~6). 
% The setup comprises a broadband supercontinuum laser, polarizing and focusing optics, a photoelastic modulator, and a photodetector to measure the polarization state of the reflected light from samples. The Kerr rotation and Kerr ellipticity were simultaneously recorded using lock-in amplification of the modulated signal at 50 kHz and 100 kHz.
The samples were positioned between the poles of an electromagnet generating a magnetic field of $B=\pm 1.3$~T, perpendicular to the measured surface. This field alternately reversed the spontaneous magnetization of the nanocones along their axes. The obtained hysteresis curves can be found in Supplementary Section~7.
The incident light was directed normally onto the metasurface, passing through a hole in the electromagnet.
In the measurements of the Tellegen effect, magneto-optical data were acquired under an external magnetic field of $B=1.3$~T. 
This ensures that the meta-atoms in all three metasurfaces remain in the same uniformly saturated magnetization state, making the metasurfaces magneto-optically equivalent and rendering our extraction technique accurate. 
The results for the measurements of the Tellegen effect in the absence of an external field are available in Supplementary Section~7.

% The measured remanent magnetization from the Kerr ellipticity and Kerr rotation hysteresis loops, shown in \textcolor{blue}{Fig. 4}, \textcolor{red}{confirms the high spontaneous magnetization of the cobalt nanocones, reaching 95\% of the saturation magnetization without an external magnetic field. This observation indicates a nearly ideal single-domain state in the cobalt segment of nanoparticles}.
% The complex MOKE, expressed as \( \theta + i\varepsilon \), is directly proportional to the ratio of the cross-polarized to the co-polarized reflected light.
% By measuring \( \theta \) and \( \varepsilon \) for the three aforementioned samples with the magneto-optical setup, the value of \( R_{\mathrm{cr}} \) is determined for each sample. Subsequently, the gyrotropic and Tellegen polarizabilities, demonstrated in \textcolor{blue}{Fig.4}, are extracted using Equation~\r{Rcr}. 

\begin{figure*}[tb]
	\centering
	\includegraphics[width=1\linewidth]{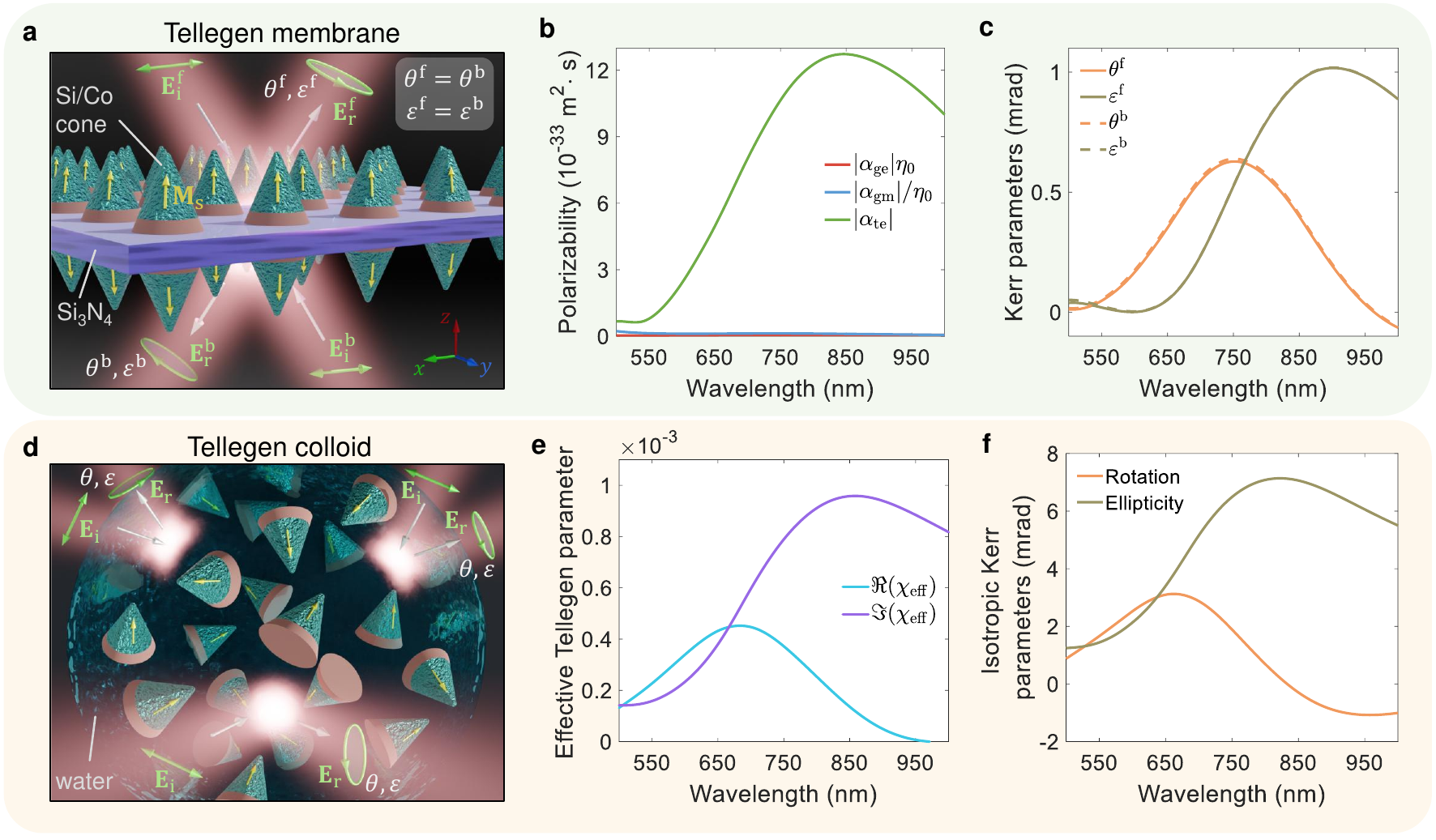} 
	\caption{\textbf{Simulated results for purely Tellegen material systems.} (a) A thin (50~nm) silicon nitride membrane with two identical Tellegen metasurfaces shown in Fig.~\ref{fig1}(a) being deposited on its both sides. The membrane is illuminated from opposite directions by normally incident light (an oblique incidence is depicted for visual clarity).  (b) Simulated complex collective polarizabilities of the membrane in (a), normalized by the free-space impedance $\eta_0$. (c) Kerr rotation angle and Kerr ellipticity for the membrane in (a) calculated for forward and backward light illuminations. (d) Isotropic Tellegen colloid consisting of randomly oriented nanocones inside water. The colloid exhibits the same magneto-optical Kerr effect regardless of the illumination direction. (e) Effective bulk Tellegen parameter of the colloid. (f) Kerr rotation angle and Kerr ellipticity for the colloid in (d).
        }
	\label{fig4}
\end{figure*}
Figures~\ref{fig3}(d)--(f) present the experimental Kerr rotation and ellipticity for the three metasurfaces ($\theta_{1,2,3}$ and $\epsilon_{1,2,3}$) as functions of the free-space wavelength. For comparison, we also include the simulated Kerr parameters for periodic metasurfaces. It is important to note that these full-wave simulations cannot fully capture the complexity of the experimental metasurfaces, featuring meta-atoms with significant variations in their overall sizes (although with the same proportions) and quasi-random distances between neighboring meta-atoms. 
In the simulations, we assume a periodic arrangement of meta-atoms with a period of $a = 330$~nm, using averaged geometric parameters of the meta-atoms depicted in the inset of Fig.~\ref{fig3}b). 

Despite the inherent differences between the quasi-random experimental metasurfaces and their periodic simulated counterparts, we observe both qualitative and satisfactory quantitative agreement between the measured and simulated Kerr spectra. All the spectra exhibit resonant dispersion. 
For the first metasurface with a spacer thickness of $h_1$, the simulated Kerr parameters are slightly larger in amplitude than the measured ones, although the spectral correspondence remains strong. For the second metasurface with thickness $h_2$, the experimental and simulated data show good agreement both spectrally and quantitatively. However, for the third metasurface with $h_3$, the simulated results are redshifted by approximately 150~nm relative to the experimental data.

By applying our proposed extraction technique (summarized in Fig.~\ref{fig2}), we calculate the polarizabilities of the experimental metasurfaces using both measured and simulated data from Figs.~\ref{fig3}(d)--(f). The extracted polarizability spectra are plotted in Figs.~\ref{fig3}(g)--(i). One can observe a good qualitative agreement between the measured and simulated data, both spectrally and in terms of amplitude. All three effects—gyroelectric, gyromagnetic, and Tellegen—exhibit comparable amplitudes within the wavelength range of interest. 
Comparing these results with the simulated polarizabilities of the idealized meta-atom in Fig.~\ref{fig1}, we observe similar response, although the resonance frequencies of the effects have shifted. These quantitative differences stem from the substantial geometric variations between the idealized meta-atom in Fig.~\ref{fig1}(a) and the realistic meta-atom in Fig.~\ref{fig3}(b). Furthermore, in our extraction, we model the array of meta-atoms as a homogenized sheet of electric and magnetic currents with deeply subwavelength thickness. This introduces an additional source of inaccuracy, as the actual meta-atom height is on the order of $\lambda/6$. Nevertheless, both the simulated and measured results provide strong evidence for the resonant nature of the diagonal Tellegen effect in the proposed optical metasurface. 
The magnitudes of the Tellegen polarizability $|\alpha_{\rm te}|$, extracted from both the measured data and the simulations for the idealized meta-atoms in Fig.~\ref{fig1} at 830~nm, reach approximately $6\times 10^{-33}$ ${\rm m}^2 \cdot {\rm s}$. 
These results demonstrate that the observed amplitude of the Tellegen effect is comparable to that of the resonant gyroelectric effect present in the metasurface. This implies that the Tellegen effect is sufficiently strong to be relevant for magneto-optical applications. 
The measured Tellegen polarizability  $|\alpha_{\rm te}|$ for the metasurface in the absence of an external field (in the remanent magnetization state) is around half of that for the saturated sample (see Supplementary Section~7).
This reduction arises because the remanent magnetization does not reach complete saturation: the cobalt layer exhibits a polycrystalline and slightly irregular microstructure resulting from the deposition process, which limits the achievable uniformity of magnetic alignment. A higher remanent magnetization could potentially be achieved by improving the crystallinity of the cobalt segment, for instance, through a post-deposition magnetic annealing step.
In the following section, we calculate and discuss the magnitude of the observed value and determine the corresponding bulk magnetoelectric parameter.

% \blue{Just to add here a compact paragraph from the section below on what this value means in terms of Tellegen effect for the physics of the system and for possible practical applications}

% A comparison of the simulated and experimental results for Kerr rotation and Kerr ellipticity demonstrates consistency in both their order of magnitude and their characteristic variations across the studied range. Similarly, this agreement extends to the experimental and simulated results for the gyroelectric, gyromagnetic, and Tellegen polarizabilities. \textcolor{red}{Maybe some other sentences should be added here for the details of curves (e.g. resonances).} Furthermore, a comparison of the gyrotropic and Tellegen polarizabilities obtained from experiments and simulations for realistic nanocones (\textcolor{blue}{Fig. 4}) with those derived from the idealized case (\textcolor{blue}{Fig.~\ref{fig1}}) indicates a similar order of magnitude for parameters while differences due to the effects of the surrounding medium and the different shape of cones. \textcolor{red}{Maybe some other sentences should be added here for a more detailed comparison.}

% \subsection*{Towards purely Tellegen material systems}\label{pureTellegen}
\section{Discussion}

The demonstrated Tellegen metasurface can be further used for the implementation of purely Tellegen metamaterials. In this section, we numerically explore the physics and observable effects in two conceptual material systems that can be built based on the proposed Tellegen metasurfaces: a Tellegen membrane and an isotropic Tellegen colloid, depicted in Figs.~\ref{fig4}(a) and (d), respectively. The former consists of two proposed metasurfaces deposited on the top and bottom sides of a nanoscale-thin silicon nitride membrane, designed to have spontaneous magnetizations in opposite directions, as shown in Fig.~\ref{fig4}(a). While this configuration requires challenging fabrication steps and sophisticated magnetic field sequences, it can be realized over small footprints~\cite{Nanoscale.2022}. The second material system, an isotropic Tellegen colloid, consists of randomly oriented nanocones dispersed within a host medium such as water~\cite{Nat.Commun.Safaei2024}. Both the membrane and colloid share the key property of exhibiting a purely Tellegen response, without gyroelectric, gyromagnetic, and other bianisotropic effects. 

To avoid limitations imposed by specific nanofabrication technologies, we consider idealized nanocones, as depicted in the inset of Fig.~\ref{fig1}(a), for both material systems in Figs.~\ref{fig4}(a) and (d). We also assume that the nanocones remain in an ideal single-domain state in the absence of an external magnetic field. Using full-wave simulations, we determine the collective polarizabilities $\alpha_{\rm ge}$, $\alpha_{\rm gm}$, and $\alpha_{\rm te}$ for the Tellegen membrane (Supplementary Section~3). As shown in Fig.~\ref{fig4}(b), the membrane exhibits only the Tellegen response. 
Indeed, due to the symmetry of the unit cell, the gyroelectric and gyromagnetic effects are nearly completely suppressed~\cite{PhysRevB.108.115101}. The Tellegen polarizability $\alpha_{\rm te}$ in the membrane is approximately twice larger than in the single metasurface (see Fig.~\ref{fig1}(d)) since the unit cell includes two cones instead of one.
In terms of observable effects, this membrane demonstrates the so-called isotropic Kerr effect~\cite{Nat.Commun.Safaei2024}, implying that, regardless of the illumination direction, the Kerr magneto-optical parameters exhibit the same sign, i.e., $\theta^{\rm f}=\theta^{\rm b}$ and $\epsilon^{\rm f}=\epsilon^{\rm b}$ (here, the Kerr rotation direction is defined with respect to the illumination direction). These parameters are plotted in Fig.~\ref{fig4}(c). This property stands in sharp contrast to conventional gyroelectric or gyromagnetic metasurfaces, where the Kerr parameters have opposite signs for opposite illumination directions. Moreover, the Tellegen membrane exhibits no Faraday effect.
For comparison, in Supplementary Section~8, we consider the same membrane geometry but where the magnetization in nanocones on both sides is the same. In that case, the symmetry results in a zero Tellegen effect. 

To evaluate the macroscopic effective-medium parameters of the purely Tellegen colloid shown in Fig.~\ref{fig4}(d), we apply the Maxwell-Garnett mixing rule~\cite[Sec.~7.2.2]{serdyukov_electromagnetics_2001}. Here, we assume the host medium to be water, however, other dielectric media would yield comparable results, differing mainly in the effective permittivity of the colloid. The volumetric concentration of the meta-atoms is set to $N = 3.47 \times 10^{20}~{\rm m}^{-3}$, corresponding to an experimentally feasible volume fraction of the cones $N_V = 0.1$. 
For our calculations, we use numerically extracted individual polarizabilities (which are independent of incident angle) of the idealized nanocone depicted in the inset of Fig.~\ref{fig1}(a). It is worth noting that the dipolar approximation assumed in this work is fully adequate for determining the bulk Tellegen parameter $\chi_{\rm eff}$. This is because higher-order multipoles, even if present in individual meta-atoms, do not contribute to this parameter when the meta-atoms are randomly distributed in a dilute colloid~\cite{barron_molecular_2009}.

Figure~\ref{fig4}(e) presents the effective bulk calculated Tellegen parameter, $\chi_{\rm eff}$, exhibiting a Lorentzian-type resonant spectral shape. At resonance near 850~nm, $\chi_{\rm eff}$ reaches approximately $10^{-3}$, which is at least 100 times larger than values observed in natural materials such as chromium oxide~\cite{B.B.Krichevtsov_1993}. 

We can further make a rough estimate of the magnitude of $\chi_{\rm eff}$ using experimental data from Fig.~\ref{fig3}(i). Since the experimentally extracted collective Tellegen polarizability $\alpha_{\rm te}$ in Fig.~\ref{fig3}(i) is of the same order as the simulated individual Tellegen polarizability~\cite{niemi_polarization_2012}, the resulting bulk Tellegen parameter $\chi_{\rm eff}$ would also be of the order of $10^{-3}$ ($\chi_{\rm eff}$ is linearly proportional to the Tellegen polarizability in the first-order approximation~\cite[Eq.~(7.108)]{serdyukov_electromagnetics_2001}).
% To put in quantitative perspective, one can also make an estimation of the order of magnitude of the bulk Tellegen parameter using previously obtained experimental data from Fig.~\ref{fig3}(i). Here, we replace the simulated individual Tellegen polarizability by the experimentally extracted collective Tellegen polarizability $\alpha_{\rm te}$, while keeping the electric and magnetic polarizabilities from the full-wave simulations. The resulting value of the Tellegen bulk parameter reaches $\chi_{\rm eff}=8 \times 10^{-4}$ at the resonance. Although it is a rough estimation, it shows that the order of magnitude
% \blue{Since $\chi_{\rm eff}$ is a linear function of the Tellegen polarizability, and the collective and individual polarizabilities have the same orders of magnitude in dilute meta-atom concentrations, we can
%

Owing to its isotropic response, the Tellegen colloid exhibits identical magneto-optical behavior regardless of the illumination direction. Figure~\ref{fig4}(f) plots these ``isotropic'' Kerr parameters for the colloid, showing qualitative similarity to those of the membrane in Fig.~\ref{fig4}(c), as both systems exhibit a purely Tellegen response. The estimated Kerr parameters are also 2 orders of magnitude higher than those in natural Tellegen materials such as chromium oxide~\cite{B.B.Krichevtsov_1993}. Indeed, the Kerr rotation and ellipticity depend linearly on the $\chi_{\rm eff}$ parameter~\cite[Eq.~(3.143)]{lindell_electromagnetic_1994}, and therefore, have the same level of enhancement.

% \blue{Our assumption is justified for the case of randomly oriented nanocones without significant agglomeration. Even if some degree of agglomeration occurs within the colloid, it would not critically affect the isotropic Tellegen response due to the nature of this effect. This is because the Tellegen effect in the present system does not rely on bulk light propagation through the medium. Instead, it predominantly originates at the vacuum--matter interface and manifests as cross-polarization upon reflection. Consequently, the dominant optical interaction responsible for the measurable Tellegen response occurs at the surface of the colloid rather than through deep penetration into its volume.}

%The twofold difference in magnitude arises from the higher nanocone density in the colloid (Fig.~\ref{fig4}(d)) compared to the membrane (Fig.~\ref{fig4}(a)).
%The presented ME metamaterial in this study exhibits remarkable potential beyond natural materials. 
% without going into practical aspects of specific nanofabrication approach, we consider how the idealized nanocones ...

In conclusion, we have demonstrated the first experimental realization of the optical Tellegen effect in an artificial metamaterial system, with the strength of around 100 times greater than that of natural materials. The Tellegen metasurface was produced using conventional materials and a well-established, highly-scalable nanofabrication, ensuring practicality for diverse applications. 
Since the presented optical metasurface also exhibits gyroelectric and gyromagnetic effects, we introduced an approach for independently extracting each of the three nonreciprocal effects. This technique is expected to be instrumental in the magneto-optical characterization of other nonreciprocal optical metasurfaces, particularly where symmetry permits multiple nonreciprocal effects. Experimental measurements, including those of magneto-optical Kerr parameters and nonreciprocal polarizabilities, demonstrate strong agreement with full-wave simulations, confirming the accuracy and reliability of the proposed method.

The inherent scalability of the chosen bottom-up nanofabrication approach and strong remanent magnetization inside meta-atoms facilitate the straightforward extension of the proposed nonreciprocal optical Tellegen metasurface to other metamaterial systems, including bulk Tellegen metamaterials and colloids.
We expect that with a minor modification of the fabrication process (introducing an additional polymer layer beneath the nanocones that can facilitate lift-off process for the nanocones into water solution), it would be possible to realize the Tellegen colloid. Such an approach would benefit from the scalability of hole-mask colloidal lithography, which enables the preparation of multiple cm-scale samples suitable for producing colloids with high meta-atom concentrations. The possible agglomeration of the magnetic nanocones inside the colloid can be avoided using a standard procedure applying static electric charge to the colloid to make the meta-atoms repel one another~\cite{tschope_nanoscale_2014a}.
Furthermore, by tailoring the geometry and composition of meta-atoms, the resonant wavelength can be precisely tuned across the optical spectrum. Opening-up the ability to modulate the Tellegen metamaterial in space and/or time has profound implications for generating and analyzing effective axion fields— which are inherently nonuniform~\cite{marsh_proposal_2019,nenno_axion_2020,guo_light_2023}— and other related phenomena~\cite{wilczek_two_1987,qi_inducing_2009,li_dynamical_2010,tercas_axion-plasmon_2018,seidov_hybridization_2023}.

% Notably, the meta-atoms are designed to possess spontaneous magnetization, eliminating the need for an external magnetic field to be applied to the metasurface.

% The magnetoelectric or Tellegen metamaterial holds considerable importance for fundamental physics, particularly quantum field theory, as it provides new possibilities for the experimental realization of various concepts related to axion matter, including dyon quasiparticles~\cite{qi_inducing_2009}, anyon statistics~\cite{qi_inducing_2009}, the Witten effect~\cite{wilczek_two_1987}, and axionic polaritons~\cite{li_dynamical_2010, tercas_axion-plasmon_2018}. Furthermore, the reported results could lead to novel developments and applications in the optical regime, including nonreciprocal light reflection and transmission~\cite{prudencio_optical_2016}, photonic topological edge states~\cite{he_photonic_2016}, and directional dichroism~\cite{rikken_observation_1997-2, train_strong_2008}, which have remained elusive from a practical standpoint. 

\section{Methods}

% \subsection*{Meta-atom design}\label{meta-atom design}
% The initial design phase commenced with a cobalt nanocylinder featuring an aspect ratio of 1, with both height and diameter set to 96 nm to sustain a single-domain state~\cite{Nat.Commun.Safaei2024}. However, due to the limitations of the HCL nanofabrication technique, the geometry was modified from cylindrical to conical, with an increased aspect ratio to enhance shape anisotropy, a crucial factor in stabilizing the single-domain state. The apex angle of the cone was precisely set at $35.5^\circ$, based on SEM measurements of the initially fabricated metasurface comprising cobalt nanocones. Considering the established cone angle, both the base diameter and silicon thickness were changed to achieve resonance of the Tellegen response within the visible spectrum. Due to the stronger shape anisotropy of the cone compared to the cylinder, combined with its high aspect ratio exceeding 1, the final design is ensured to maintain a single-domain state.

\subsection{Simulation}\label{simulation}
The full-wave simulation characterization of the metasurfaces was performed using COMSOL Multiphysics software with a frequency-domain solver. Periodic boundary conditions were applied to simulate an infinite domain around the structure, with the polarization of the incident light defined at the incident port. The complex reflection spectra were then calculated based on the scattering parameters. We used the material dispersion properties of the alumina, aluminum, cobalt, and amorphous silicon from measurements reported in \cite{alumina}, \cite{Mirshafieyan:16}, \cite{Co.diagonal,Co.offdiagonal}, and \cite{Si}, respectively. To account for the inevitable oxidation of the bare aluminum layer in the first fabricated metasurface sample (which has a target alumina thickness of $h_1=0$~nm), we incorporated an alumina oxidation layer with a thickness of 9~nm in our full-wave simulations used to generate the results in Fig.~\ref{fig3}(d).

\subsection{Sample fabrication}
\label{Sample fabrication}

To achieve large-scale fabrication of optical Tellegen metasurfaces, we employed hole-mask colloidal lithography~\cite{HCL}. Microscope slide glass substrates (VWR International) with dimensions of 24~mm × 24~mm × 0.4~mm were used. The substrates were cleaned through an ultrasonication process in acetone. Subsequently, a 5~nm titanium (Ti) adhesion layer and a 100 nm aluminum (Al) layer were deposited using electron-beam physical vapor deposition (EB-PVD). Atomic layer deposition (ALD) was utilized to deposit aluminum oxide (Al$_2$O$_3$) films with thicknesses of 60~nm and 120~nm for the second and third metasurfaces, respectively.
A polymethyl methacrylate (PMMA) layer (950PMMA, 4\% solution in anisole, MW = 950,000, MicroChem) with a thickness of 250 nm was spin-coated on all metasurfaces and soft-baked at 130°C for 3 minutes on a hot plate. Following the baking process, the PMMA layer was subjected to oxygen plasma treatment for 15 seconds to enhance its hydrophilicity. The PMMA-coated substrate was then functionalized with polydiallyldimethylammonium chloride (PDDA, 0.2\% solution in deionized (DI) water), followed by rinsing with DI water and drying with a nitrogen (N$_2$) gun.
Negatively charged polystyrene beads (100~nm, sulfate latex, 0.2\% suspension in DI water, Invitrogen) were deposited onto the substrate via pipetting and left in contact for 120 seconds. After careful rinsing with DI water, the samples were dried using an N$_2$ gun at normal flow incidence to prevent bead aggregation. A 10 nm chromium (Cr) layer was subsequently deposited, followed by tape-stripping of the polystyrene spheres (SWT-10 tape, Nitto Scandinavia AB). Reactive ion etching (RIE) with oxygen plasma was then performed to etch through the entire PMMA layer, forming an evaporation mask.
The silicon (Si) and cobalt (Co) layers were deposited using EB-PVD. During material deposition, a progressive reduction in the nanohole diameter occurred as a result of the accumulation of material along the hole edges, leading to a decrease in the diameter of the nanodiscs and the formation of conical meta-atoms. The final step involved the removal of the PMMA layer via acetone lift-off.

%This technique leverages the self-assembly of negatively charged sulfate latex microspheres, facilitating the cm-scale fabrication of metasurfaces.The fabrication process begins with the deposition of Ti/Al and Al$_2$O$_3$  layer with the further spin-coating of polymethyl methacrylate (PMMA) layers. Next, polystyrene beads are sedimented, followed by the deposition of a chromium layer, which serves as a mask. Subsequent tape-stripping removes the beads, and oxygen plasma etching is applied through the entire thickness of the PMMA layer to form the hole-mask. Finally, silicon and cobalt are deposited, and the PMMA layer is lifted off with acetone. During material evaporation, a progressive reduction in the nanohole diameter occurs due to the deposition of material on the hole edges. This shrinkage results in a corresponding decrease in the diameter of the nanodiscs, ultimately leading to the formation of conical meta-atoms. While this effect is negligible for low-aspect-ratio meta-atoms, such as nanodiscs, it becomes significant when the deposited material thickness approaches the feature diameter~\cite{HCL}. A detailed description of the fabrication process can be found in Supplementary Section~8.

\subsection{Magneto-optical characterization}\label{MOKE}  
The magneto-optical Kerr effect (MOKE) of the samples was characterized using an NKT SuperK EXW-12 supercontinuum laser connected to an acousto-optical tunable filter (AOTF), providing a wavelength range of 500–1400 nm. The AOTF employed in our setup has two outputs: VIS/NIR (500–900 nm) and NIR2 (800–1400 nm). A focusing doublet, Glan-Thompson polarizer, and silver D-shaped mirror were used to direct the light into the magnet poles and focus it onto the samples, which were positioned between the poles of a 3470 GMW electromagnet (Fig.~\ref{fig3}(c)). The measurements were performed at normal incidence. The reflected beam from the samples passed sequentially through a photoelastic modulator I/FS50 Hinds Instruments (PEM), a half-wave plate Thorlabs FR600HM (HWP), and a beam splitter Thorlabs WPA10 (BS), before being collected by a balanced photodetector Thorlabs PDB250A2 (BPD). In the MOKE setup, the combination of the photoelastic modulator with two lock-in amplifiers operating at 50 kHz (Signaloc 2100, Hinds Instruments) and 100 kHz (SR830 DSP, Stanford Research Systems), along with a balance detector, enabled the simultaneous measurement of Kerr rotation ($\theta$) and Kerr ellipticity ($\epsilon$).
To ensure reproducibility, we fabricated and characterized 3 samples for each Al$_2$O$_3$ thickness. Additionally, measurements were conducted at several regions on each sample. No significant differences in the optical response of the metasurfaces were observed, confirming good reproducibility of both the fabrication and measurement processes.

\section*{Acknowledgments}
% \textcolor{red}{Note: all authors, please complete or update this section for your group.}
S.S.J. and V.A.  acknowledge the Academy of Finland (Project No. 356797). S.S.J. acknowledges the Walter Ahlström Foundation and the Finnish Foundation for Technology Promotion. V.A.  acknowledges the PREIN Flagship (Project No. 320167) and the Finnish Foundation for Technology Promotion. S. F. acknowledges a MURI project from the U. S. Army Research Office (Grant No. W911NF-24-1-0224). I.F., R.C., and A.D. acknowledge the Swedish Research Council for Sustainable Development (Formas) (Project No. 2021-01390), Thuréus Forskarhem och Naturminne Foundation, and the Swedish Research Council (VR) (Project No. 2024-05025). N.K. acknowledges financial support from the Magnus Ehrnrooth Foundation. We acknowledge the Micronova Nanofabrication Centre, supported by Aalto University, for providing resources for this work.  We also thank Julius Hohlfeld and Junyoung Hyun for their assistance with the measurement setup.

\section*{Competing interests}
The authors declare no competing interests.

\section*{Availability of data and materials}
The authors declare that the data supporting the findings of this study are available within the paper, in the Supplementary file, and are available from the corresponding authors upon request.

\section*{Author contributions}
V.A. conceived the idea and determined the optimal topology of the meta-atom. S.S.J. and V.A. conducted the theoretical calculations. S.S.J. and I.F. performed the full-wave simulations of empirical samples. I.F. and S.S.J. were responsible for the fabrication. S.S.J. and R.C. conducted the experimental measurements and N.K. provided assistance and support during the measurement process. S.S.J., V.A, and I.F. characterized the experimental and simulated results.
S.v.D., S.F., A.D., and V.A. supervised the work. S.S.J. drafted the manuscript, with all authors contributing to its review and revision. All authors contributed to the discussions of the results.

\section*{References}
\bibliography{references}

\end{document}